\documentclass[a4paper,11pt]{amsart}
\usepackage[latin1]{inputenc}
\usepackage{amsmath, amsthm, amssymb, amsfonts, amscd}
\usepackage[english]{babel}
\usepackage[all]{xy}
\usepackage[linktocpage=true]{hyperref}
\usepackage{color}
\usepackage{pinlabel}
\usepackage{tikz-cd}
\usepackage{cite}
\usepackage{enumitem}

\newtheorem*{thm*}{Definition}
\newtheorem*{thm2*}{Theorem}

\theoremstyle{remark}

\numberwithin{equation}{section}
\newcommand\restr[2]{{
  \left.\kern-\nulldelimiterspace 
  #1 
  \vphantom{\big|} 
  \right|_{#2} 
  }}

\title[\tiny Grid diagrams to investigate knot spaces and topo II simplification]{Grid diagrams as tools to investigate knot spaces and topoisomerase-mediated simplification of DNA topology}

  \author[A. Barbensi, D. Celoria, H. A. Harrington, A. Stasiak, D. Buck]{Agnese Barbensi, Daniele Celoria, Heather A. Harrington, Andrzej Stasiak, Dorothy Buck}
\address{AB, DC, HAH: Mathematical Institute, University of Oxford, Oxford, UK.}

\address{AS: Center for Integrative Genomics, University of Lausanne, Lausanne, Switzerland and SIB Swiss Institute of Bioinformatics, Lausanne, Switzerland.}
\address{DB: Department of Mathematical Sciences, University of Bath, Bath, UK and Mathematics/Biology, Trinity College of Arts \& Sciences, Duke University, Durham, NC, USA.}

\begin{document}

\begin{abstract}
  Grid diagrams with their relatively simple mathematical formalism provide a convenient way to generate and model projections of various knots. It has been an open question whether these 2D diagrams can be used to model a complex 3D process such as the topoisomerase-mediated preferential unknotting of DNA molecules. We model here topoisomerase-mediated passages of double-stranded DNA segments \\through each other using the formalism of grid diagrams. We show that this grid diagram-based modelling approach captures the essence of the preferential unknotting mechanism, based on topoisomerase selectivity of hooked DNA juxtapositions as the sites of intersegmental passages. We show that grid diagram-based approach provide an important, new and computationally convenient framework for investigating entanglement in biopolymers. 
\end{abstract}

\maketitle
\section*{Introduction}

DNA topology is regulated by enzymes called topoisomerases. A class of these, known as type II topoisomerases \cite{(1)}, act on double stranded DNA molecules by introducing double-stranded DNA breaks that are bridged by the bound enzyme \cite{(2)}. Subsequently, a distinct dsDNA segment (either from the same molecule or from another one) is
passed through and the cut is resealed \cite{(1)}. Since topoisomerases are of vital importance for the proper functioning of DNA replication \cite{(3)} and of several other cellular processes \cite{(4)}, they are often used as targets for antibacterial and anticancer drugs \cite{(5),(6)}. Although type II topoisomerase's ability to cleave and reseal DNA molecules plays such a fundamental role, performing inter-segmental passages on long polymers often results in the creation of non-trivial knots and catenanes (also known as \emph{links}). DNA knots are known to be damaging for the cell \cite{(7)}. Therefore, they should be quickly unknotted by DNA topoisomerases. Recent determination of human chromosome's structure from single cell HI-C data revealed that chromosomes can be knotted as well \cite{(8)}. Remarkably, it has been observed in reactions performed in vitro that when type II topoisomerases act on randomly cyclised DNA molecules (\emph{i.e.}~molecules having the equilibrium level of knotting), the level of knotting decreases dramatically \cite{(9)}. Thus, type II topoisomerases manifest a preference to unknot the DNA, and many biological explanations for this behaviour have been proposed \cite{(9), (10), (11), (12), (13)}.  Rybenkov \emph{et al.}, in their paper establishing the concept of preferential unknotting by DNA topoisomerases, proposed that type II topoisomerases form clamps that actively slide along the DNA, concentrating DNA entanglements and thus facilitating DNA unknotting \cite{(9)}. Since the active sliding mechanism was not confirmed experimentally, the same group hypothesised later that topoisomerase creates a sharp bend in the DNA region that will be transiently cut during the reaction and that this directs the passage of the transported segment through the transient cut by passing from inside to outside of the bend formed by the DNA topoisomerase \cite{(10)}. More recently, it was hypothesised that type II topoisomerases are preferentially unknotting DNA due to their ability to specifically recognise and act on hooked juxtapositions of DNA segments \cite{(13)}. Lattice based simulations \cite{(14),(15)} and equilateral chain model simulations \cite{(16)} have confirmed this hypothesis. 
 
However, these prior simulations, and previous investigations \cite{(17),(18)} of connectivity between neighbouring knot spaces via single intersegmental passages, necessitated the use of computationally expensive randomisation algorithms to ensure uniformity of sampling and to produce systems exhibiting global and detailed balance, while undergoing a given type of intersegmental passages (as discussed in the results). Hence our aim here is to propose a new, computationally convenient, purely topological and intrinsically randomised framework to examine how the interconversion rates between different knot types depend on the local geometry of regions where the intersegmental passages occur. 

Since all knots can be represented by planar projections encoding the information of over/under passing segments (\emph{i.e.}~knot diagrams), we use planar diagrams and more specifically grid diagrams \cite{(19), (20)} for our analysis. We provide exact enumeration of the segment passages between grid diagrams taken up to a certain complexity called the \emph{grid number} (GN). This measure of complexity for a knot diagram can be easily related to the length of the polymer associated to the diagram; grids with grid number $n$ can be thought of polymers with $2n$ statistical segments \cite{(21)}. Our approach enables us to compute exact values for the distribution of knot types after an intersegmental passage between diagrams with GN less than 8. In the case of higher complexity diagrams the complete enumeration is impractical, so we randomly sample the space of diagrams and we determine the topological consequences of intersegmental passages.
     
We first apply the grid diagram approach to estimate the probabilities of passing from one knot type to another through the action of a hypothetical unbiased topoisomerase acting on circular DNA molecules of various lengths. We then test the consequences of passages occurring at ``hooked juxtaposition'' \cite{(13)} and calculate the resulting knotting reduction factor.
\vspace{0.8cm}

\textbf{Acknowledgements:} The authors thank Marc Lackenby and Cristian Micheletti for helpful and insightful conversations. A.B. and H.A.H are supported by the RS-EPSRC grant ``Algebraic and topological approaches for genomic data in molecular biology'' EP/R005125/1. H.A.H gratefully acknowledges funding from a Royal Society University Research Fellowship and the Oxford return carers fund. D.C. has received support from the European Research Council (ERC) under the European Union's Horizon 2020 research and innovation program (grant agreement No 674978). D.B. and A.S acknowledge support from the Leverhulme Trust, Grant RP2013-K-017. We also thank the COST Action European Topology Interdisciplinary Action (EUTOPIA) CA17139 for supporting collaborative meeting of the authors.

\vspace{0.8cm}
\textbf{Author contributions:}
AB, DB and DC  conceived and designed the research. AB and DC performed the research. AB, DB, AS, HAH wrote the article and discussed the results. All authors read and commented on the article.

\vspace{0.8cm}

\textbf{Data availability:} The data is available on the GitHub repository: https://github.com/agnesedaniele/DNAandGrids

\section*{Results}

\subsection*{Grid diagrams}
\begin{figure}[h]
\includegraphics[width=8cm]{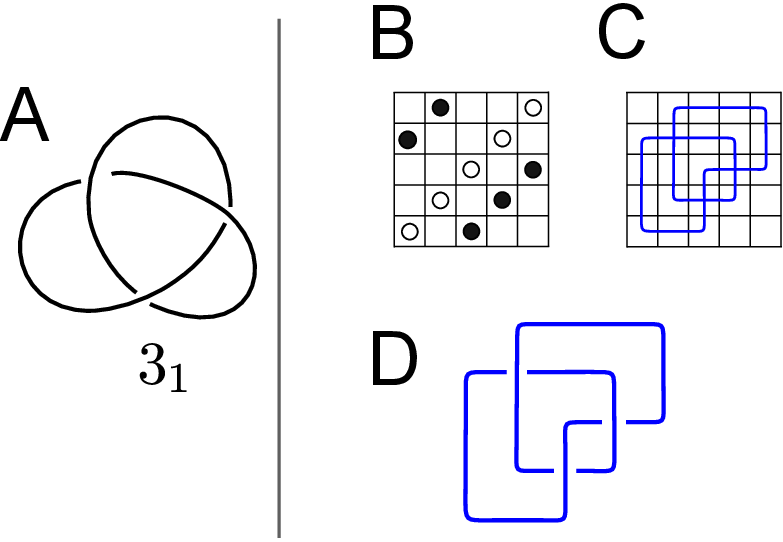}
\caption{\textbf{A standard knot diagram and a grid diagram of a knot.} \textbf{(A)} Standard diagram of a left-handed trefoil knot. We use Alexander-Briggs' notation of knots, where the first number indicates the minimal crossing number of a given knot type, and the subscript number denotes its tabular position amongst all knots with that crossing number. Standard knot diagrams are scale-free and therefore do not inform about the number of statistical segments of a knotted polymer they represent. \textbf{(B-D)} Generation of a grid diagram of a trefoil knot. Grid diagram formalism requires that the square grid with n rows and n columns has exactly one segment in each row and column of the represented polygonal chain. A grid diagram of a knot configuration is generated in 3 steps. \textbf{(B)} Place $2n$ markings (dots) corresponding to ends of the modelled polygonal chain segments. Markings are placed following a ``Sudoku'' rule, requiring that each column and each row of the grid contains exactly two markings in distinct squares. \textbf{(C)} Each pair of markings in the same row/column is connected by a segment. \textbf{(D)} For the segments that intersect, we follow the convention that the vertical segment passes over the horizontal segment.}
\label{fig:1}
\end{figure}

Grid diagrams are a special kind of knot diagrams, first introduced in \cite{(19)} and widely used in knot theory (see \emph{e.g}~\cite{(20)}). Grid diagrams consist of a square, planar, $n \times n$ grid ($n$ is a natural number greater than 2), in which $2n$ markings are placed, corresponding to vertices of a piecewise linear curve representing a given knot. Each row/column of the grid must contain exactly two markings (see Figure \ref{fig:1}, \textbf{(B)}), and two markings can not occupy the same position. A knot diagram can be created from a grid diagram as follows: connect by a segment any two markings on the same row or column, and impose every vertical strand to be over-passing. We remark here that the overpassing condition is not restrictive, since any knotted configuration can be represented by a grid diagram \cite{(20)}. Just as for classical knot diagrams \cite{(22)}, there exists a finite set of moves that allows one to represent any deformation of the underlying curve (the knotted DNA molecule, in our case) in terms of local transformations of the grid (\emph{i.e.}~local displacement of the markings \cite{(20)}). Refer to the Supplementary Information in Appendix \ref{Appendix} for more details on grid diagrams.
 
There is an immediate advantage when considering grid diagrams, as opposed to other ways of modelling knots (\emph{e.g.}~classical diagrams, lattice or stick models): each grid diagram can be efficiently described by a pair of permutations (that is, a pair of $n$-tuples) on $n$ elements, determining the positions of the markings. The number $n$ is a measure of complexity for grid diagrams, and it is called the \emph{grid number} GN. A grid diagram of size $n$ can be seen as corresponding to an equilateral polymer chain with $2n$ statistical segments, see \emph{e.g.}~\cite{(21)}. For example, grid diagrams in GN 5 can be used to estimate the knotting probability of a polymer having the length of $\sim 10$ statistical segments, which in the case of DNA molecules would correspond to $\sim 3$-kb. 
     
\begin{figure}[h]
\includegraphics[width=12cm]{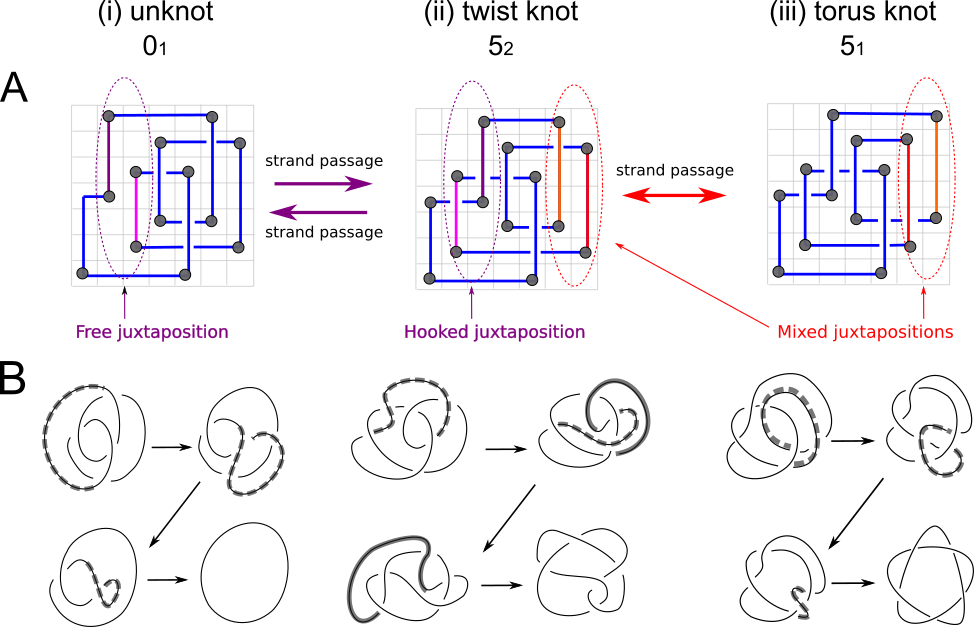}
\caption{\textbf{Interleaving commutations as a model for strand passages.} \textbf{(A)} Intersegmental passages between grid diagrams representing the $(i)$ unknot $0_1$, the $(ii)$ twist knot $5_2$ and the $(iii)$ torus knot $5_1$. Intersegmental passages are achieved by interleaving commutation of two consecutive rows/columns such that the interiors of their corresponding intervals intersect non-trivially, but neither is contained in the other. The exchanged segments are highlighted in red and purple. When the area delimited by two parts of the diagram between two consecutive crossings forms a rectangle a hooked juxtaposition is formed. A juxtaposition can be ``strongly hooked'' if the rectangle is a square. The diagram of $5_2$ contains one strongly hooked-juxtaposition. Performing an interleaving commutation at that hooked juxtaposition transforms the $5_2$ into the trivial knot. This exchange transforms the hooked juxtaposition into a ``free'' one. Juxtapositions which are neither hooked nor free are called mixed. Performing an interleaving commutation at the highlighted mixed juxtaposition on $5_2$ transforms this knot into the $5_1$, shown right. \textbf{(B)} The local deformations that transform the three grid diagrams in \textbf{(A)} into their respective standard diagrams. Thick grey lines highlight the arcs involved in the various deformations.}
\label{fig:2}
\end{figure}

Local deformations and strand passages are then realised by changing the pair of permutations. In particular, strand passages (\emph{i.e.}~single actions of type II topoisomerases) are achieved on grids by a process called \emph{interleaving commutation} which exchanges the positions of two adjacent and interleaved rows or columns, as described in Figure \ref{fig:2}, \textbf{(A)}. This operation does not increase the complexity of the grid diagram, so can be easily encoded. This combinatorial description of grid diagrams allows us to perform exact enumeration for a range of complexity and, theoretically, it allows computations with grids of arbitrary dimensions.

\subsection*{Free and hooked juxtapositions.}
The part of a diagram between consecutive crossings can take different shapes, some of which correspond to the projection of a hooked juxtaposition. We call a \emph{hooked juxtaposition} the part of a grid diagram in which the segments between two consecutive crossings geometrically form a rectangle (see Figure \ref{fig:2}, $(ii)$). In analogy with the lattice models \cite{(4),(15)}, we test the effects of geometrical selection of sites (where the intersegmental passage happens) on the topological outcome (see Figure \ref{fig:2}). By imposing that the crossing changes happen only at the specific local configurations resembling the hooked geometries \cite{(16)}, we can test the hooked-juxtaposition hypothesis \cite{(13)} that type II topoisomerases achieve disentanglement by performing strand passages only at hooked juxtapositions. We measure how much a juxtaposition is hooked using the length of the maximum segment of such rectangle as a parameter: this quantification of hooked means that the larger the parameter value, the less the configuration is ``hooked''. We call ``strongly hooked'' the juxtapositions in which the rectangle is the elementary square. Note that, as shown in Figure \ref{fig:2}, a strand passage occurring at a hooked juxtaposition transforms the geometric site into a ``free'' juxtaposition. 

\subsection*{Knot interconversion fluxes resulting from unbiased intersegmental passages are balanced for all realisable configurations in grids with GN 6.}
\begin{figure}[h]
\includegraphics[width=12cm]{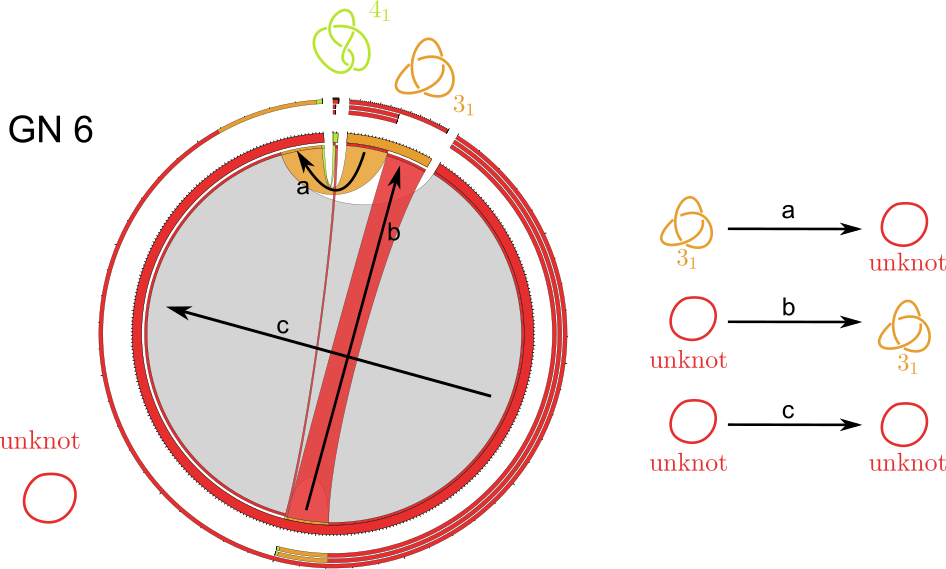}
\caption{\textbf{Visualisation of strand passage-mediated knot-interconversion fluxes using circos plots.} The three layers of thin external arcs, progressing from inside to outside, represent outgoing, incoming and total fluxes involving a given knot type, respectively. These external arcs are segmented to indicate how the respective fluxes were redistributed. The thickness of the (interior) chords connecting different knot types reflects the fraction of outgoing and incoming knot interconversion fluxes between given types of knots. The chords representing knot interconversion fluxes are coloured as the knot type these fluxes originate from, with the exception of chords starting and ending in the same knot type, that are in grey. These correspond to the fluxes resulting from strand passages not changing the knot type. The bases of the chords representing outgoing fluxes are coloured according to the knot type a given flux leads to. The bases of chords representing incoming fluxes are left white. The length of the various thick arcs around the circumference, coloured as the corresponding knot diagrams, indicate the sum of fluxes outgoing from and incoming to a given knot type. The global observed flux in a given system is normalised to 1, which corresponds to the circumference of the circle. }
\label{fig:3}
\end{figure}

The circos plot in Figure \ref{fig:3} summarizes the data obtained in calculations where all configurations of knots (including trivial knots) that are realisable as grid diagrams with GN 6 undergo intersegmental passages resulting from unbiased interleaving commutations. The thickness of cords connecting arcs representing different knots is proportional to the interconversion fluxes connecting these knots. Each pair of knots $i$ and $j$ is connected by two cords representing the interconversions of all the configurations representing the knot $i$ into configurations representing the knot $j$ and interconversions of the knot $j$ into the knot $i$. As can be seen, these incoming and outgoing cords connecting a given pair of knots have the same thickness at their starting and ending portions. Thus, for example, the cord representing the outgoing flux from trefoil knots to trivial knots (marked with an arrow \textbf{a}) has the same thickness as the cord representing outgoing flux from trivial knots to trefoil knots (marked with an arrow \textbf{b}). The circos plots we used convey even more information about the knot interconversion fluxes and the legend to Figure \ref{fig:3} explains how this information is encoded.

\subsection*{The configuration space of grid diagrams with GN$\leq 7$. }
\begin{figure}[h]
\includegraphics[width=12cm]{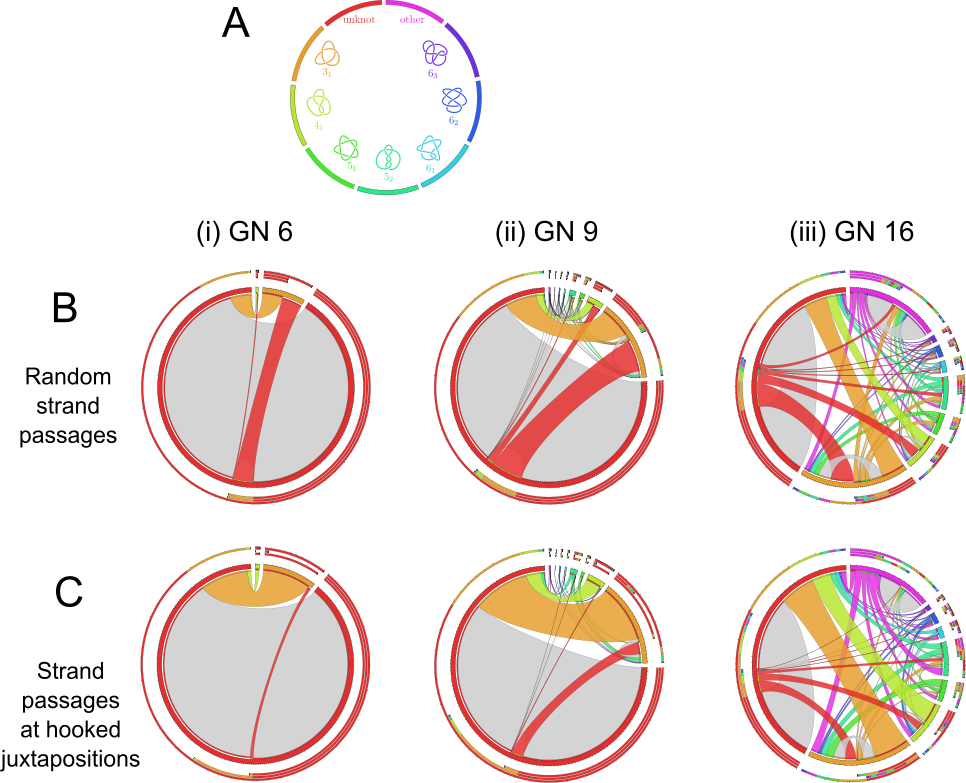}
\caption{\textbf{Knot interconversions occurring at hooked juxtapositions lead to preferential unknotting.} \textbf{(A)} Colour guide informing which colours correspond to which knots in circos plots shown in \textbf{(B)} and \textbf{(C)}. A comparison of the topological consequences of unbiased strand passages \textbf{(B)} with the ones resulting from strand passages occurring only at strongly hooked juxtapositions \textbf{(C)}. Comparison of circos plots for small grid diagrams $(i)$ with these for larger grid diagrams $(ii)$,$(iii)$ shows that as the system gets more complex, more types of knots are formed and they contribute stronger to the knot interconversion fluxes. In the circos plots representing fluxes resulting from unbiased strand passages \textbf{(A)} the incoming and outgoing fluxes connecting any pair of knots are of the same intensity. This indicates that the generated set of grid diagrams represents the topological equilibrium. When the same set of grid diagrams undergoes intersegmental passages involving only strongly hooked juxtaposition the interconversion-fluxes from the trefoil to the unknot are much more intense than the opposite fluxes. This effect is especially strong for smaller GNs.  }
\label{fig:4}
\end{figure}

There are a total of 1,859,118 different grid configurations with grid number GN$\leq 7$, of which 1,773,114 are unknotted, 78,296 are (left or right-handed) trefoils, 6014 are figure eight knots, 798 are $5_1$ torus knots and 882 are $5_2$ twist knots, and only 14 of them are $8_{19}$ knots. As an example, the unbiased adjacency table of knots with GN$=6$ is visualised in the left-most circos plot on Figure \ref{fig:4}, \textbf{(B)}, $(i)$. We observe that, in agreement with previous works (17,18), most ($\sim 91.7\%$) of the strand passages occurring in unknotted diagrams do not change the topology of these diagrams, and, out of those passages that change the topology, $\sim 94.5\%$ transform the unknot into the trefoil.
The circos plots summarising the sampling of GN 9 and 16 are shown in Figure \ref{fig:4}, \textbf{(B-C)}, $(ii)$-$(iii)$. It is immediately apparent how the knot-type fauna becomes more variegated as the GN increases, with considerably higher occurrence of complex knots, and strand passages-mediated fluxes becoming more visible. In the exact enumeration of unbiased strand passages, the number of diagrams passing from the $i$-th knot type to the $j$-th knot type is equal to those passing from the $j$-th to the $i$-th. Thus, the fact that outgoing and incoming fluxes are equal shows that performing all the interleaving commutations does not introduce any selection bias. In higher GNs, we have to ensure that the sampling method enables us to describe the system at equilibrium. As mentioned in the Introduction and in the Material and Methods section, by performing every strand passage on each one of the configurations randomly sampled, we achieve detailed and global balance effortlessly. This can be seen from the circos plots of Figure \ref{fig:4}, \textbf{(B)}, $(ii)$-$(iii)$, in which the sizes of the arcs representing incoming and outgoing fluxes of every pair of knot types correspond almost exactly.

\subsection*{Evolution of the connectivity between knot spaces.}
\begin{figure}[h]
\includegraphics[width=12cm]{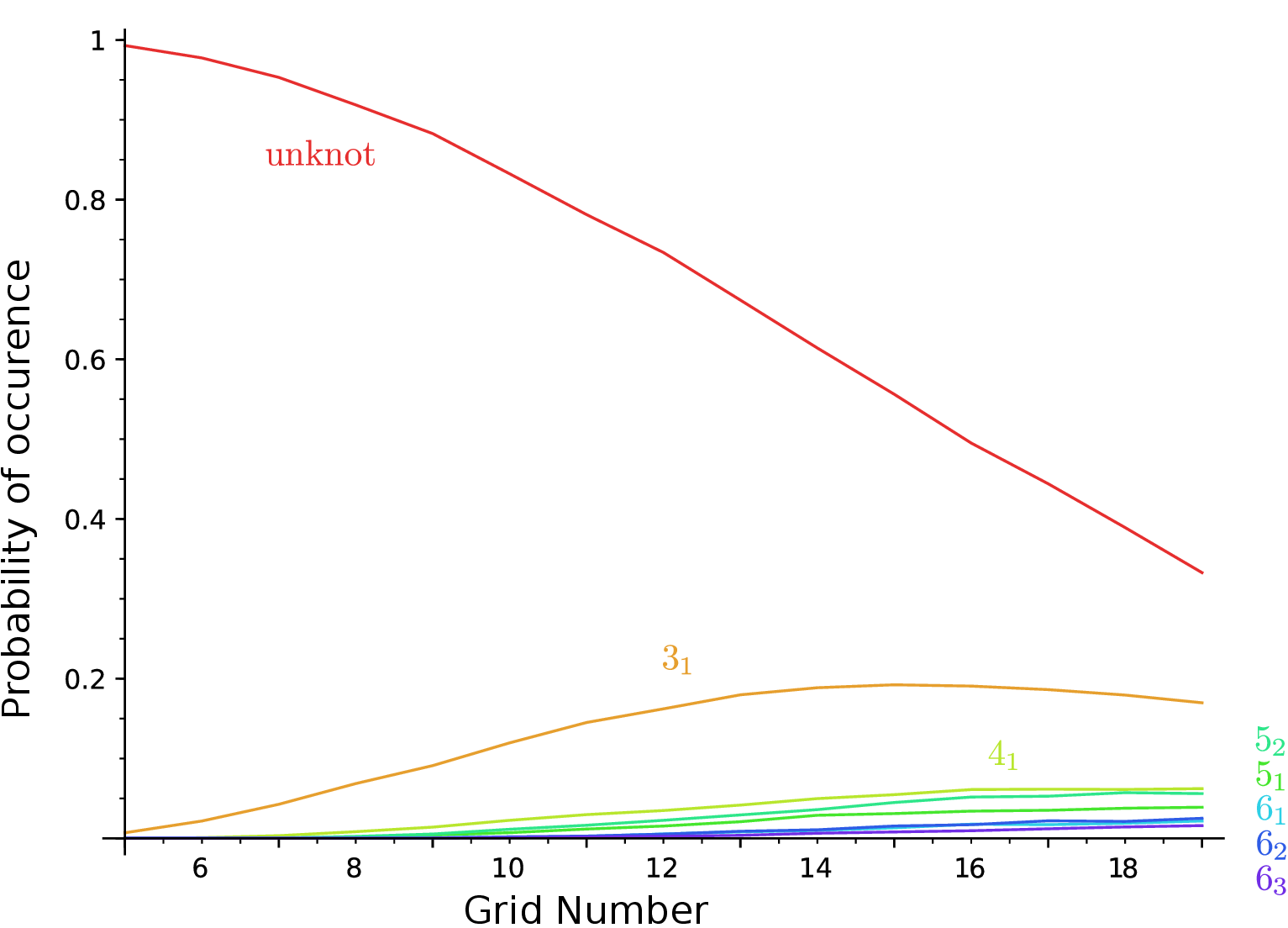}
\caption{\textbf{Occurrence probability of prime knots as a function of GN.} The curves give the occurrence probabilities of knots with minimal crossing number up to 6, plotted as a function of the grid number. The occurrence probability of a given knot type is defined as the ratio between the configurations representing that knot type and the total number of configurations.   }
\label{fig:5}
\end{figure}

It is well known \cite{(23),(24)}, that for closed polymers, the probability that a configuration is unknotted decreases as the length of the polymer increases. The same behaviour can be observed for grid diagrams. Figure \ref{fig:4} and Figure \ref{fig:5} show how the configuration spaces of a knotted molecule evolve as the complexity given by the GN increases. The probability that a configuration of a given knot type is converted into another knot type is called the \emph{transition probability} of the first knot type towards the second. Note that as GN increases, the probability of occurrence of unknotted conformations decreases monotonically and that the transition probabilities towards complex knot types increases monotonically. For example, the probability of passing from an unknotted diagram to a trefoil knot passes from an initial value of  $\sim 4.3\%$ in the exact computation for GN$=5$, to the $19.64\%$ in complexity 20. Unsurprisingly, for every GN, the transition probability for the unknot passing to a non-trivial knot is consistently the highest for the trefoil.

One may be tempted to \emph{sort} knot types connected through a single strand passage to another knot, in terms of the amount of observed strand passages towards the knot types in question, as discussed in \cite{(17)} (there, they refer to this concept as ``interface area'' between knot spaces). In our setting, we can formalise the heuristic notion of \emph{knot closeness} between two given knot types as the ratio between the occurrences of intersegmental passages leading to the interconversion between the two knot types and the total number of intersegmental passages that lead to interconversion between any two knot types.
As an example, the data discussed above suggests that the trefoil is the closest knot type to the trivial knot. The observation that such properties are maintained as the GN increases, coupled with the fact that our results are topological in nature (hence independent of any geometric or physical feature of the model), indicates that these are intrinsic features of the various knot types. Figure \ref{fig:5} shows how the knot fractions change as the GN increases. 

We observe that the trend is qualitatively similar to the case of equilibrated polygonal chains in 3D (see \emph{e.g.}~Figure $2$ of \cite{(26)}).  However, in polygonal chains formed by essentially 2D grid diagrams, knotting is stimulated as compared to 3D situation \cite{(26)}. In our case, the population of non trivial knots starts to exceed $50\%$ at GN 17, and every knot with minimal crossing number $\leq 13$ can be represented by grids of size  $\leq 12$ \cite{(20)}. This highlights a further advantage of our framework. Namely, we are able to account for complex knot types while working with configurations of relatively small size.

\subsection*{Topological simplification through geometric selection of sites and the knotting reduction factor.}
\begin{figure}[h]
\includegraphics[width=12cm]{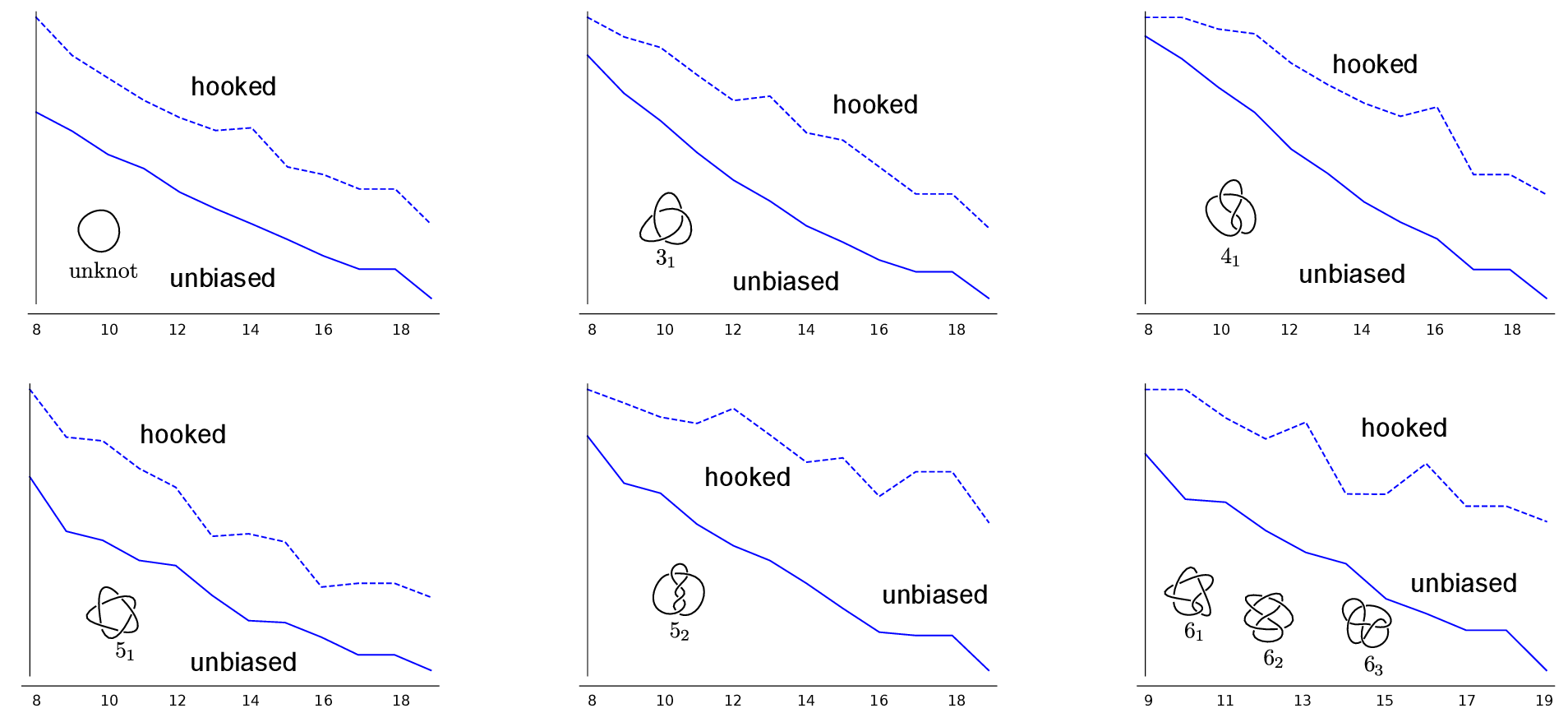}
\caption{\textbf{Topological simplification due to hooked juxtapositions.} The plots show the transition probabilities of each knot type towards ``simpler'' knots, as a function of the grid number. In each plot, the dotted line refers to strand passages happening at strongly hooked juxtaposition, while for the other we consider unbiased interleaving commutations. In the case of the unknot, the $3_1$ and the $4_1$ we consider only the unknotting probabilities. For the $5_1$ and the $5_2$ we consider passages towards the $3_1$, the $4_1$ and the unknot. Finally, for 6 crossings knots, we plot the transition probabilities towards knots with lower ``length over diameter ratio'' \cite{(31)} (thus, for the $6_1$ we consider only passages towards knots with crossing number less than or equal to 5, while for the $6_2$ and the $6_3$ we consider also passages towards the $6_1$ and towards the $6_1$ and $6_2$ respectively).}
\label{fig:6}
\end{figure}

Figures \ref{fig:4} and \ref{fig:6} show how the knot interconversion fluxes resulting from intersegmental passages occurring once at every interleaved juxtaposition are affected by limiting the intersegmental passages to interleaved juxtapositions that are strongly hooked. In the case of passages occurring once at every interleaved juxtaposition (Figure \ref{fig:4}, \textbf{(B)}) the fluxes between any two knot types are equilibrated. This is best seen for small GNs, where we enumerate all possible configurations. In these cases the number of observed interconversions from trivial knots into trefoil knots, for example, was exactly the same as the number of interconversions that changed trefoil knots into trivial knots. Importantly, the probability that a given configuration of a trivial knot inter-converted into a trivial knot was, for the analysed grid sizes, always substantially smaller than the probability that a given configuration of a trefoil knot is converted into trivial knot. However, the number of unknotted configurations realisable was larger than the one of trefoil knots. The difference in the numbers of unknots and trefoils was such that the sum of all interconversions from unknots into trefoils was exactly the same as the sum of all interconversions from trefoils into unknots. 

The circos plots presented in Figure \ref{fig:4}, \textbf{(B)} correspond to the situation in which DNA circles with a given size are permitted to undergo random intersegmental passages till they reach an equilibrium. Such a system at equilibrium shows both global balance and detailed balance. Recall that global balance means that the fraction of DNA molecules forming any given knot type reaches its equilibrium level and will not show a tendency to increase or decrease that level, although one may observe some fluctuations. Detailed balance means that the interconversion fluxes between any two types of knots are the same in both directions when observed over a sufficiently long time. Our modelled system, where all intersegmental passages realisable by interleaved commutations are allowed, shows a detailed balance of interconversion fluxes. Thus, it is evident that our system is at the equilibrium state.

Figure \ref{fig:4} shows what happens when the equilibrated set of configurations obtained after unbiased intersegmental passages is allowed to undergo further intersegmental passages, but this time only occurring at strongly hooked juxtapositions. The generated interconversion fluxes are no longer balanced and the fluxes from more complex knots toward simpler knots are stronger than the opposite fluxes. The ratio of these initial interconversion fluxes connecting trivial knots with trefoil knots describes how much the ratio of the number of trefoil knots and unknots would need to be diminished so that the system will reach its new equilibrium. The extent of this knotting diminution, also known as knot reduction factor, was measured experimentally for intersegmental passages mediated by the type II topoisomerase, Topoisomerase IV \cite{(9)} and for several simulated systems \cite{(14),(15),(16)}. In experimental studies by Rybenkov \emph{et al.}~\cite{(9)} it was observed that upon many rounds of type II topoisomerase-mediated intersegmental passages, the amount of trefoil-forming DNA molecules was greatly reduced as compared to the amount that would be obtained after random intersegmental passages. The experimentally observed reduction of trefoil knot concentration was approximately 90-fold for 7-kb DNA circles and approximately 50-fold for 10-kb DNA circles \cite{(9)}. The knotted population seen in the experiments by Rybenkov \emph{et al.}~\cite{(9)} for 10-kb DNA circles divides into approximately $3\%$ trefoils and $97\%$ unknots. A similar ratio in our model can be found in the range $5\leq$GN$\leq7$. We therefore analyse all knot diagrams realisable in grids of those sizes in order to determine what is the knotting reduction factor resulting from passages occurring only at strongly hooked juxtapositions.

We began with the enumeration of passages occurring at all juxtapositions and observed that there were 6240 interconversions from configurations representing the trefoil knot to configurations representing the unknot and the same number of interconversions from unknots to trefoil knots. We then analysed passages at hooked juxtapositions and observed that there were 1220 interconversions from configurations representing the trefoil knot to configurations representing the unknot and only 80 interconversions from unknots to trefoil knots. Therefore, to equilibrate the interconversion fluxes observed when passages are limited to strongly hooked juxtapositions, the proportion of trefoil knots to unknots would need to be diminished over 15 times as compared to situation where the regions of passages are not selected. This would happen after many rounds of passages limited only to hooked-intersegmental juxtapositions. Note that the 15-fold knotting reduction factor concluded from our simulations is strong, but it is still significantly smaller than 50-90 fold knotting reduction factor observed experimentally \cite{(9)}. 

Earlier simulation studies by Burnier \emph{et al.}~\cite{(16)} showed that the more a juxtaposition is hooked, the larger the corresponding knotting reduction factor. In that study, a 23-fold knotting reduction was observed for hooked juxtapositions with acute angles smaller than 25$^\circ$, whereas in our model, we are limited to right angles between segments forming hooked parts of the polymer. It is possible that the effective bending in hooked juxtapositions interacting with the type II topoisomerase, Topoisomerase IV, is stronger than the one we can model using grid diagrams. Our model can nevertheless capture the principle of knotting reduction resulting from geometrical selection of sites for intersegmental passages.

In experiments performed by Rybenkov \emph{et al.}~\cite{(9)} 7-kb long DNA circles showed a knotting reduction factor nearly two times higher than 10-kb long DNA circles. The inverse relation between the knotting reduction factor and the size of circular DNA molecules stems naturally from the mechanism proposed in \cite{(13)}, in which a higher ratio between the bending rigidity and the size of DNA molecules corresponds to a higher type II topoisomerase-mediated knot reduction factor. This effect was confirmed in simulation studies using polygonal knots in the 3D lattice model \cite{(14),(15)}.

As our model is essentially a 2D model it is of interest to analyse how knot reduction factor changes with the grid size. Remarkably, for grid diagrams with GN$=5$ (where there are 10 diagrams forming the trefoil knot) the knotting reduction factor connected to intersegmental passages at hooked juxtapositions was infinite. All 10 diagrams of the trefoil knots had hooked juxtapositions and all passages occurring at these juxtapositions converted the trefoil knot to the unknot. On the other hand, all passages occurring at strongly hooked juxtapositions in unknotted diagrams did not result in a change of topology. As already mentioned, for GN$=6$ the knotting reduction factor was approximately 15. Analysis of diagrams in GN$=7$ showed that the selection of sites results in a knotting reduction factor of about 8. Therefore, we can conclude that planar grid diagrams capture adequately the experimental observation that knotting reduction by type II topoisomerases is strongest when acting on short DNA molecules \cite{(9)}.

\section*{Discussion}
Grid diagrams provide a new tool to investigate knot adjacency and the configuration space of knots. They provide a way to examine the statistical and probabilistic properties of knotted polymers without the need to rely on initial configurations as starting points. Thanks to their intrinsic combinatorial definition, grid diagrams allow us to uniformly sample the space of simple and even complex knots in arbitrary large grid diagrams. This is in contrast with analysis in the equilateral chain model and lattice model, for which uniformity of sampling for long polymers is computationally expensive.
Further, since complex knot types (\emph{i.e.}~with high minimal crossing number) arise in relatively low grid number with significant percentages, this allows us to easily investigate statistical features of knotted polymers in a varied population of knots.
For these reasons, we believe that the grid diagram approach provides a complementary and valid reference system helping to better understand mechanisms of action of various DNA topoisomerases, strand passage events and even polymers in general.

\section*{Materials and Methods}

\subsection*{Modelling of dsDNA molecules and of type II topoisomerase-mediated actions using knot diagrams.}
Although DNA forms a double helix, for such topological aspects as DNA knots, DNA helicity can be frequently neglected and the formed knot can be recognised from a projection of the axial path of analysed DNA molecules \cite{(1)}. Such projections can be conveniently represented as grid diagrams \cite{(20)} (see Figures \ref{fig:1} and \ref{fig:2}). Grid diagrams encode the information of which arc is over-passing and which is under-passing at each crossing (see Figure \ref{fig:1}, \textbf{(A)}). A single action of a type II topoisomerase corresponds to performing a \emph{crossing change} in the diagram (that is, exchanging the over and under passing arcs), as described in \cite{(25)}. It is well-known that any knot type admits infinitely many different diagrams \cite{(22)}, so in this setting we can model the configuration space of a circular DNA molecule undergoing the action of a type II topoisomerase as a directed network. In this network, the vertices are the 2D projections of the knotted configurations (\emph{i.e.}~the diagrams), and these vertices are connected through directed edges, each representing a single crossing change (\emph{i.e.}~an action of type II topoisomerase). A \emph{subspace} of the network is formed by those vertices that correspond to diagrams representing a same knot type, and two knot types whose corresponding subspaces are connected by directed edges are called \emph{adjacent.} More details on this network are available in the Supplementary Information in Appendix \ref{Appendix}.

\subsection*{The configuration space as a network.}
In our investigation we consider grid diagrams with complexity between 5 (the minimal GN in which non-trivial knots appear) and 20. We create the network of these grid diagrams where two grids of the same GN are connected by a directed edge if it is possible to transform the first diagram into the second via a single strand passage (\emph{i.e.}~an interleaving commutation). For each GN the network of grid diagrams has finitely many vertices and edges. The strand passages-mediated flux (or knot interconversion flux) from a knot type to another is the union of directed edges in the network going from the subspace corresponding to the first knot type into the subspace of the second one. The intensity of the flux is proportional to the number of these directed interconversion edges. We perform our analysis by enumerating knot conformations and by counting unbiased and ``hooked'' strand passages between conformations of given knot types. Imposing that the strand passages happen only at hooked juxtapositions changes the shape of the network of configurations. The subspaces corresponding to simple knot types become more preferable than the others, as we discussed in the result sections. Since the GN correlates with the length of the underlying closed polymer, the configuration space changes as GN increases in value (see Figure \ref{fig:5}), which is similar to models that take length as parameter \cite{(26)}. We remark that our model, being purely topological, does not consider such physical quantities such temperature. In other words, the model is temperature-independent, and each conformation is assigned the same statistical weight.

\subsection*{Exact grid enumeration.}
We begin our analysis by enumerating all grid diagrams with grid number $5\leq$GN$\leq 7$. This is achieved using a Sage \cite{(27)} program, by listing all possible pairs of permutations of length $5\leq n \leq7$, and keeping the ones representing knot diagrams. The topological state (\emph{i.e.}~the underlying knot type) of every configuration is determined using a combination of knot invariants (knot polynomials, determinant, signature \cite{(22)}). We then perform every possible crossing change from each configuration. We summarise the data in a square table, the \emph{unbiased adjacency table}, in which the $(i,j)$th-entry represents the number of strand passages from the $i$-th knot type to the $j$-th knot type. The adjacency table is then visualised using circos plots (see Figure \ref{fig:3}) \cite{(28)}. The previous process is then repeated, this time by allowing only strand passages to occur at hooked juxtapositions (see Figure \ref{fig:2}). We summarise the data in another square table, the \emph{``hooked'' adjacency table}. While the unbiased adjacency table is symmetric, the ``hooked'' one presents a strong unknotting preference, as discussed in the results section.

\subsection*{Sampling.}
Listing all possible configurations before and after strand passages provides information about adjacency between different knot types. The number of configurations grows super-factorially with GN (there are  $1/2 (n((n-1)!))^2$ different configurations in GN$=n$, see \cite{(29)}), and the number of admissible crossing changes increases more than linearly with the complexity. Thus, an exhaustive computation for higher values of GN becomes quickly infeasible, and so the investigation on diagrams with higher grid number is performed through random sampling. Since a grid diagram is completely determined by a pair of permutations defining the positions of the markings, uniformity of sampling is automatically built in our model. In fact, we only rely on the effectiveness of Python's extensively tested \texttt{random()} function \cite{(30)}. We consider all knots whose minimal crossing number is less or equal than 8 (labelling the more complex ones collectively as ``other''), since these are the most commonly identified DNA topoisomers. We randomly sample configurations of GN$=8$, and then perform every admissible strand passage (achieved by interleaving commutations, see Figure \ref{fig:2}, \textbf{(A)}). After computing the starting and resulting knot types, we obtain an unbiased adjacency table, as before. We then compare the unbiased sampled table with the hooked sampled table obtained by restricting to strand passages happening at hooked juxtapositions. The same process is repeated for every grid number, for $8\leq$GN$<20$.

\bibliographystyle{amsplain}

\newpage
\appendix
\section{Supplementary Information}
\subsection*{The network of configurations}\label{Appendix}

\subsubsection*{\textbf{Knots and their diagrams}}
A \emph{knot} is a smooth embedding of the unit circle $S^1$ into the 3-dimensional Euclidean space. We will call knots also the images of such embeddings. Thus, a knot is a closed curve in space. Two knots $K_1$ and $K_2$ are \emph{equivalent} if their corresponding embeddings are related by an \emph{ambient isotopy} (for a definition see \cite{(22)}).  This means that two knots $K_1$ and $K_2$ are equivalent if and only if the first curve can be smoothly (\emph{i.e.}~without performing cuts) deformed into the second one. A knot is called \emph{trivial} (we will refer to it as the unknot) if it is equivalent to the standard embedding (\emph{i.e.}~the unknotted one) of the unit circle $S^1$ into the 3-dimensional Euclidean space. 

We will often call \emph{knot types} the equivalence classes of knots under the ambient isotopy equivalence relation. Analogously, we call a \emph{link} an embedding of a disjoint union of $n$ circles into the sphere. The image of each circle is called a \emph{component} of the link, and a \emph{link type} is the equivalence class of such an embedding, up to ambient isotopy. Thus, a knot is a link with only one component.

Knots are often studied through their diagrams. A \emph{diagram} for a knot $K$  is a projection of the knot into a 2-dimensional plane or sphere, such that the only singularities are transverse double points. Each intersection point is called a \emph{crossing}, and it is endowed with the information of which strand is passing over (for more details and precise definitions see \cite{(22)}). Every knot type admits infinitely many diagrams representing it. Moreover, every knot diagram has a finite number of crossings \cite[Ch. 3]{(32)}.

Different knot types are distinguished using \emph{knot invariants.} A knot invariant is a map that associates an ``algebraic'' object (\emph{e.g.}~a number or a polynomial) to a knot, taking the same value on equivalent knots. Many simple knot invariants can be defined using diagrams. For example, the \emph{crossing number} of a knot $K$ is defined as the smallest number of crossings of any diagram of the knot.

\begin{figure}[h]
\includegraphics[width=11cm]{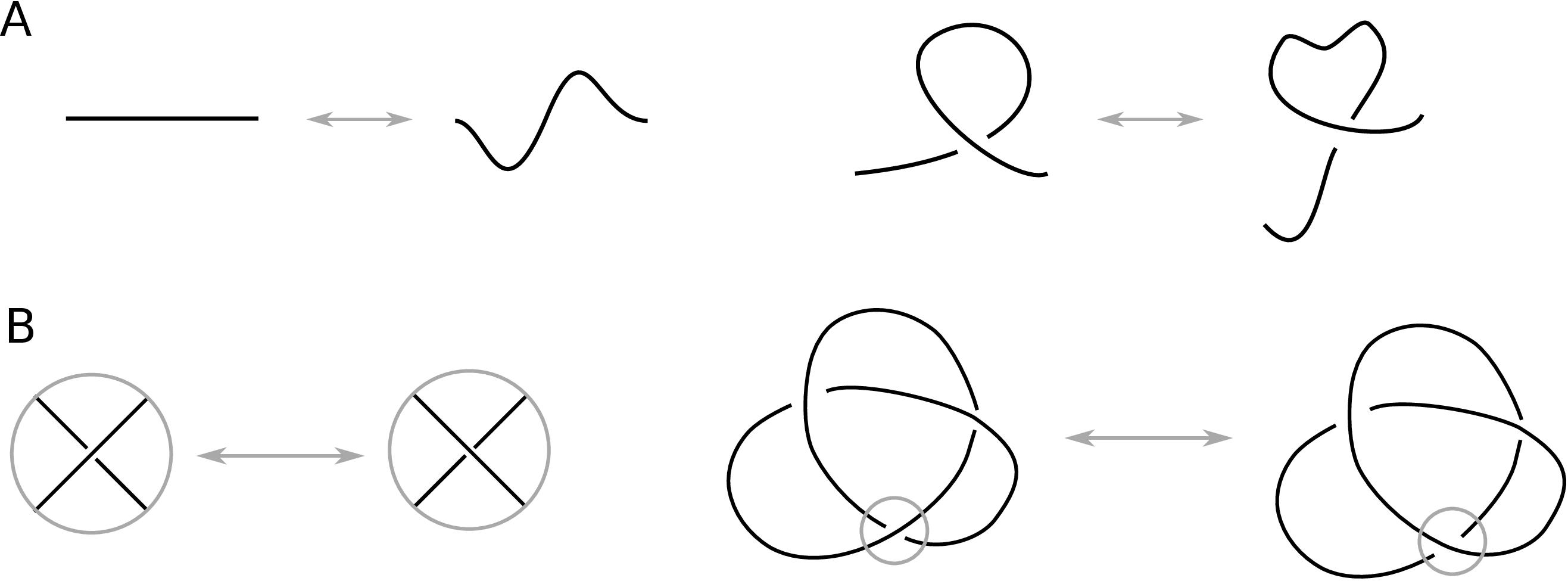}
\caption{\textbf{Local isotopy and crossing change.} \textbf{(A)} Two examples of a local planar isotopy acting on an arc. \textbf{(B)} Changing a crossing in a knot diagram. On the right, two diagrams differing by a crossing change.}
\label{fig:1s}
\end{figure}

A \emph{planar isotopy} can modify locally a knot diagram by slightly moving an arc as in Figure \ref{fig:1s}, \textbf{(A)}, or by displacing a whole diagram, without creating or removing any crossing. We consider two diagrams to be \emph{equivalent} if they differ only by a planar isotopy. We will often refer to a specific diagram as a \emph{configuration} of the corresponding knot. Strand passages between different configurations are modelled as crossing changes on the respective diagrams (see Figure \ref{fig:1s}, \textbf{(B)}). Note that performing a crossing change might result in changing the knot type represented by the diagram.

\subsubsection*{\textbf{The network of knot diagrams.}}
We can represent the set of configurations of knots as a network (\emph{cf.}~\cite{(33)}) in which:
\begin{itemize}
 \item the vertices are knot diagrams (up to planar isotopy); 
 \item the edges represent crossing changes (\emph{i.e.}~two vertices are connected by an edge if and only if they differ by a single crossing change).
\end{itemize}
	 
\begin{figure}[h]
\includegraphics[width=12cm]{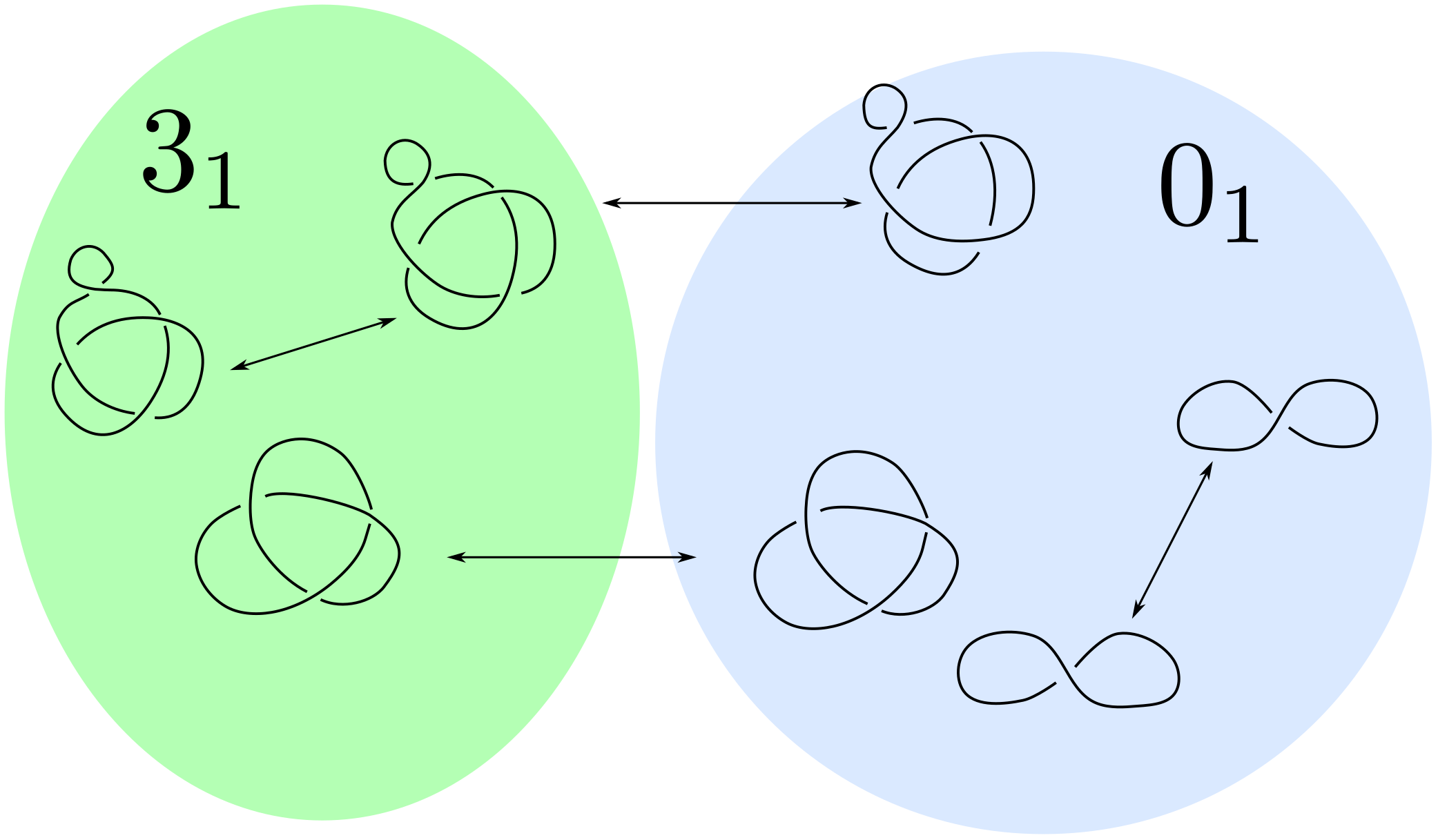}
\caption{\textbf{The network of knot diagrams.} A part of the network of configurations, consisting of 7 vertices and 4 edges. A part of the trefoil subspace (on the left) and of the trivial knot subspace (on the right) are shown. Edges represent crossing changes. Edges connecting vertices in the trefoil subspace with vertices in the trivial knot subspace form the flux between the trefoil and the unknot.   }
\label{fig:3s}
\end{figure}

There are infinitely many different knot types and for each knot type there are infinitely many different diagrams. Thus, this network is infinite. However, since each diagram has finitely many crossings, each vertex has only finitely many edges emanating from it. We show as an illustration a small part of this network in Figure \ref{fig:3s}.

The network is partitioned in knot-subspaces. Each subspace is formed by those vertices corresponding to diagrams representing the same knot type. In Figure \ref{fig:3s}, a portion of the subspaces for the trefoil knot  and the trivial knot  are shown. Given two subspaces, we call the \emph{flux} between them the union of all the edges connecting a vertex of one subspace with a vertex of the other. 

\subsubsection*{\textbf{Grid diagrams}}
Grid diagrams are a special kind of knot diagrams that provide an easy and combinatorial way to represent knots and links, first introduced in \cite{(19)} (see also \emph{e.g.} \cite[Ch.3]{(20)}, for definitions and results).

\begin{thm*}
A grid diagram is a $n \times n$ matrix $G$, together with two sets of n markings, denoted by $\mathbb{X} = \{X_0, \dots , X_{n-1}\}$ and $\mathbb{O} = \{O_0, \dots, O_{n-1}\}$. Each row and and each column of the matrix contains exactly one $X$ and one $O$ marking. A link diagram can be retrieved from a grid by connecting the $X$ with the $O$ in each row and column. In the crossings we let the vertical strand always pass over the horizontal one. The size of the matrix is a natural number $n\geq 2$, called the grid number GN of $G$.
\end{thm*}

\begin{figure}[h]
\includegraphics[width=9cm]{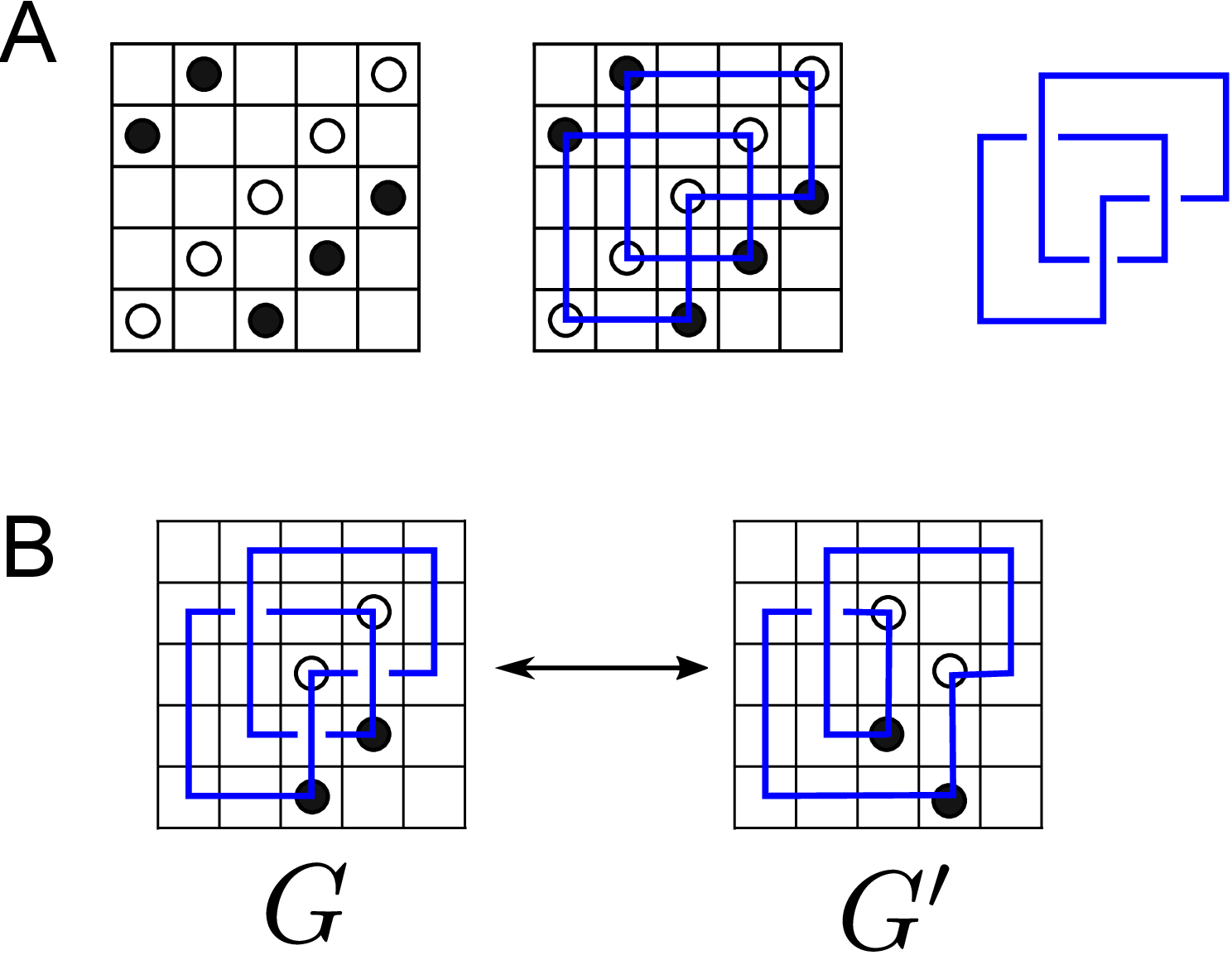}
\caption{\textbf{Grid diagrams.} \textbf{(A)} A grid diagram representing the trefoil knot. Here the black circles represent the $O$-markings and the white circles the $X$s. It follows that $\sigma_{\mathbb{X}} = (0,1,2,3,4)$ and $\sigma_{\mathbb{O}} = (3,4,0,1,2)$. \textbf{(B)} An interleaving commutation between the grid diagrams $G$ and $G'$.}
\label{fig:2s}
\end{figure}

An example of a grid diagram representing the trefoil knot is shown in Figure \ref{fig:2s}, \textbf{(A)}. A grid diagram can be described by two permutations $\sigma_{\mathbb{X}}$ and $\sigma_{\mathbb{O}}$. If there is an $O$ marking in the intersection of the $i$-th column and the $j$-th row, then $\sigma_{\mathbb{O}}$ maps $i$ to $j$. The number of disjoint cycles in which the permutation $\sigma_{\mathbb{X}} \cdot \sigma_{\mathbb{O}}^{-1}$ splits is the number of components of the corresponding link diagram (link diagrams are defined exactly as knot diagrams) \cite{(20)}. In particular, if the number of cycles is 1, the diagram represents a knot.

\begin{thm2*}\cite[Ch.3]{(20)}
Any link $L$ can be represented by a grid diagram. In this setting, strand passages (\emph{i.e.}~crossing changes) are modelled as local moves on the grid. These moves are called interleaving commutations, see Figure \ref{fig:1s}, \textbf{(B)}.
\end{thm2*}

\begin{thm*}
Fix two consecutive columns (or rows) in a grid diagram $G$, such that their corresponding intervals intersect non-trivially, but neither is contained in the other. Let $G'$ be the grid obtained by swapping these two columns (or rows). We say that the grid diagrams $G$ and $G'$ are related by an interleaving commutation.  
\end{thm*}

\begin{thm2*}\cite[Ch.3]{(20)}
If two grid diagrams $G$ and $G'$ are related by an interleaving commutation, their corresponding knot types $K$ and $K'$ are related by a crossing change. 
\end{thm2*}
 
\subsubsection*{\textbf{The networks of grid diagrams.}} 
We can repeat the ideas from the previous section using grid diagrams instead of regular diagrams. In this case, we have an infinite network whose vertices are grid diagrams, and edges represent interleaving commutations. Again, since the number of interleaving row/columns of a grid diagram is bounded by a function of the grid number, every vertex has finitely many edges emanating from it. 

Note that interleaving commutations do not change the size (the grid number GN) of a diagram. Thus, we can split the network of grid diagrams into a collection of finite components having grids with the same size as vertices. 
More precisely, call $\mathcal{G}$ the network of grid diagrams. Then $\mathcal{G}$, can be written as the disjoint union $\mathcal{G} = \bigsqcup_{n>1} \mathcal{G}_n$, where each $\mathcal{G}_n$ is the network of grid diagrams with GN$=n$, connected by edges representing interleaving commutations\footnote{Note that we start from n$= 2$ since grid diagrams are not defined for n$= 1$. However, n$= 5$ is the minimum grid number that allows grid diagrams representing non trivial knots. This is why in our computations we start from n$= 5$ instead.}.

Since for every $n$ there are finitely many possible grid diagrams having GN$=n$, each network  is finite. In particular, in each $\mathcal{G}_n$ the flux between any two subspaces is made up by finitely many edges. 

The specific nature of grid diagrams allows us to differentiate between different local geometries, as explained in the Materials and Methods of the main manuscript. Imposing that strand passages (\emph{i.e}~interleaving commutations) happen only at specific local geometries (such as hooked juxtapositions, see Material and Methods in the main manuscript) induces an orientation of the edges, hence the network is directed.
 
\subsection*{Computations and results}\label{Appendix1}
A pair of permutations on $n$-elements $\sigma_{\mathbb{X}}$ and $\sigma_{\mathbb{O}}$, represents a grid diagram for a link whenever $\sigma_{\mathbb{X}}(i) \neq \sigma_{\mathbb{O}}(i)$ for all $i = 0, \cdots ,n-1$. 
Moreover, by checking the number of cycles in $\sigma_{\mathbb{X}} \cdot \sigma_{\mathbb{O}}^{-1}$, we can tell whether it represents a grid diagram for a knot.

\subsubsection*{\textbf{Exact enumeration}}
We generate all possible pairs of permutations on $n$ elements for $n=5,6,7$, and check whether they represent a grid diagram of a knot. 

For each grid, we determine the corresponding knot type using a combination of several knot invariants (Alexander and Jones polynomials, determinant, and signature, see \cite{(22)} for definitions). We only consider knot types whose crossing number is $\leq8$, and we label the configurations representing more complex knot types as \emph{other}. We then perform all possible interleaving commutations from the grid, and compute the knot type for each of the resulting grid diagrams. 

We collect the data in a square matrix, called the \emph{(unbiased) adjacency table}. 
The $(i,j)$-entry of this table is the number of computed interleaving commutations from grids representing the knot type $i$ to grids representing the knot type $j$ (the knot types are ordered following the Alexander-Briggs' notation \cite{(22)}). Note that since we are performing an exact enumeration, this table is symmetric.
We then restrict to interleaving commutations that occur only at hooked juxtapositions (see main manuscript, Materials and Methods). Recall from the main manuscript that we measure how much a juxtaposition is hooked using the length of the maximum segment of such a rectangle as a parameter: the larger the parameter value, the less the configuration is hooked. 

Again, we collect the information in several adjacency tables for passages at hooked juxtapositions. More precisely, for each grid number GN we created GN-3 (hooked) adjacency tables. Each of these tables accounts for passages happening at juxtapositions where the length of the maximum segment of the rectangle is $\leq l$, where $1 \leq l \leq \text{GN}-3$. The probability that a configuration representing a specific knot type is transformed into an unknotted one by an interleaving commutation increases as we consider hooked juxtapositions with shorter edges.

The complete code is available at \cite{(34)}.

\subsubsection*{\textbf{Sampling}}
For each GN within the range 8-19 (refer to the main manuscript for the chosen range of grid numbers) we proceed by randomly sampling a pair of permutations, and checking if it represents a valid knot grid diagram. We collect a certain number of configurations, which depends on GN (for higher GN we performed 16 iterations of the process, each time sampling 1000 configuration; for smaller GN we iterated the process twice, sampling 8000 configurations each time). We then proceed as done for the ``Exact enumeration'' on this sample. Again we produce several adjacency tables (both unbiased and hooked). The complete code is available at \cite[Sampling.sage]{(34)}.

\subsubsection*{\textbf{Data}}
The adjacency tables are available at \cite{(34)}. Each file contains the data relative to a specific grid number GN (or range of GNs), for a total of 16000 configurations sampled in each GN. In each file, for each iteration, we write:
\begin{itemize}
 \item the GN;
 \item the total number of configurations considered (e.g.~the number of sampled configurations plus the number of configurations reached through strand passages);
 \item the number of sampled configurations; 
 \item the distribution of knot types in the sample; 
 \item the distribution of knot types with respect to the total number of configurations considered; 
 \item a list of adjacency tables, starting from the unbiased one, followed by the hooked ones. The hooked adjacency tables are ordered from less to more hooked.
\end{itemize}

\vspace{8mm}

\vspace{5cm}
\end{document}